\documentclass[aps,prd,preprint,superscriptaddress,showpacs,floatfix,nobibnotes,nofootinbib]{revtex4-2}
\usepackage[usenames, dvipsnames]{color}

\usepackage{float}

\usepackage{latexsym}
\usepackage{amsmath}
\usepackage{amssymb}
\usepackage{graphicx}
\usepackage{setspace}
\usepackage{longtable}
\usepackage{enumitem}
\usepackage[bottom]{footmisc}

\usepackage[normalem]{ulem}
\usepackage[usenames, dvipsnames]{color}

\usepackage{setspace}

\usepackage{bm}
\usepackage{epsfig}
\newcommand{\bea}{\begin{eqnarray}}
\newcommand{\eea}{\end{eqnarray}}

\newcommand{\nn}{\nonumber}

\begin{document}
\setlength{\baselineskip}{20pt}

\title{Phase transitions in 3-dimensional Dirac semi-metals using Schwinger-Dyson equations}

\author{Margaret E. Carrington}
\email[]{carrington@brandonu.ca} 
\affiliation{Department of Physics, Brandon University,
Brandon, Manitoba R7A 6A9, Canada}
\affiliation{Winnipeg Institute for Theoretical Physics, Winnipeg, Manitoba, Canada}

\author{Wade N. Cowie}
\email[]{cowiewn57@brandonu.ca}
\affiliation{Department of Physics, Brandon University,
Brandon, Manitoba R7A 6A9, Canada}
\affiliation{Winnipeg Institute for Theoretical Physics, Winnipeg, Manitoba, Canada}

\author{Brett A. Meggison}
\email[]{brett.meggison@gmail.com} 
\affiliation{Department of Physics \& Astronomy, University of Manitoba, Winnipeg, Manitoba, R3T 2N2 Canada}
\affiliation{Winnipeg Institute for Theoretical Physics, Winnipeg, Manitoba, Canada}

\date{October 19, 2023}

\begin{abstract}
We study the  semi-metal/insulator quantum phase transition in three-dimensional Dirac semi-metals by solving a set of Schwinger-Dyson equations. We study the effect of an anisotropic fermion velocity on the critical coupling of the transition. We consider the influence of several different approximations that are commonly used in the literature and show that results for the critical coupling change considerably when some of these approximations are relaxed. Most importantly, the nature of the dependence of the critical coupling on the anisotropy depends strongly on the approximations that are used for the photon polarization tensor. On the one hand, this means that calculations that include full photon dynamics are necessary to answer even the basic question of whether the critical coupling increases or decreases with anisotropy. On the other hand, our results mean that it is possible that anisotropy could provide a mechanism to promote dynamical gap generation in realistic  three-dimensional Dirac semi-metallic materials.

\end{abstract}

\maketitle
\newpage

\section{Introduction}
\label{intro-sec}

There has been much recent interest in the properties and mathematical description of three dimensional (3d) Dirac semi-metals. 
This has in part  been stimulated by the discovery of graphene which has some  similar special electronic properties (reviews can be found in \cite{neto1,neto2,gusynin07}).  
The free Hamiltonian of the 3d Dirac fermions can be written
\bea
H_0 = \int d^3\vec r\,\bar\Psi_\sigma(\vec r)
\big(v_1\gamma_1\nabla_x+v_1\gamma_2\nabla_y + v_3\gamma_3\nabla_z\big)
\Psi_\sigma(\vec r)
\eea
where $\Psi_\sigma(\vec r)$ is a four component spinor and $\vec v_F = v_1\hat i+v_1\hat j + v_3 \hat k$ is the fermion velocity which has magnitude much less than the speed of light. The index $\sigma\in(1,N)$ is the fermion flavour which has the physical value $N=2$, corresponding to the two Dirac cones in the Brillouin
zone. We work in Euclidian space and the gamma matrices are defined to satisfy the algebra $\{\gamma_\mu,\gamma_\nu\}=2\delta_{\mu\nu}$. 
The long-range Coulomb interaction between Dirac fermions is described by the Hamiltonian
\bea
H' = \frac{e^2}{4\pi \epsilon} \int d^3\vec r d^3\vec r' \bar\Psi_\sigma(\vec r) \gamma_0 \Psi(\vec r) \frac{1}{|\vec r-\vec r'|}\bar\Psi_\sigma(\vec r') \gamma_0 \Psi(\vec r')\,
\eea
where $\epsilon$ is a dielectric constant whose value is strongly material dependent \cite{neto2}.
The total Hamiltonian has a continuous chiral symmetry that is broken if a gap $\sim \langle\bar\Psi_\sigma \Psi_\sigma\rangle$ is dynamically generated. 
A discussion of  different types of 3d Dirac semi-metals can be found in \cite{nature}. 
Among many interesting applications, it has recently been proposed that 3d Dirac materials could be used to detect sub-MeV dark matter \cite{Hochberg:2017wce}.
The textbook \cite{ivan} gives an excellent overview of the electronic properties of 3d Dirac semi-metals. 

One important property of these materials is that the electron dispersion relation close to the Dirac points is linear and the material can  be described with a low energy effective field theory approach \cite{kane,gonz1,xiao2}.
The effective strength of the Coulomb interaction is represented by the parameter $\alpha = e^2/(4\pi \epsilon v_1)$. 
Typically one defines the critical coupling as the value of $\alpha$ for which the gap goes to zero. 

In typical 3d Dirac semi-metals 
the $z$-component of fermion velocity is
much smaller than the components in the $x$-$y$ plane. To quantify the anisotropy we define  $\eta=v_3/v_1$; in the isotropic limit $\eta=1$ and strongly anisotropic materials have $\eta\ll 1$. 
The main goal of this paper is to investigate the dependence of the semi-metal/insulator phase transtion on the anisotropy parameter  $\eta$. We note that phase transitions in isotropic systems have been studied in \cite{gonz1,gonz2}. 
In \cite{xiao2,xiao1} it was pointed out that when the goal is to study  the effects of anisotropy, it might not be appropriate to use the conventional definition of the critical coupling as the value of $\alpha = e^2/(4\pi \epsilon v_1)$ for which the gap goes to zero.
 Their argument is as follows. 
If $\eta=v_3/v_1$ is changed while $\alpha$ is held fixed then  $v_3$ changes at fixed $v_1$. 
This means that the total kinetic energy of the 3d Dirac fermions changes, while the strength of the Coulomb interaction does not. The point is that even though the coupling $\alpha$ is held fixed, 
the ratio of the potential to kinetic energies, which physically quantifies the relative strength of the Coulomb interaction, changes. 
The authors of \cite{xiao2,xiao1} propose that $\bar v =(v_1^2 v_3)^{1/3}$ gives a kind of mean value of the components of the fermion velocity and that one should study the dependence of the gap function on the parameter $\tilde\alpha =e^2/(4\pi \bar v) = \alpha/\eta^{1/3}$. We will consider both definitions and discuss how results depend on the definition that is used. 
 
Calculations using an effective field theory that describes non-relativistic and anisotropic fermions are complicated for several reasons. 
The Lorentz symmetry that is present in QED is broken by the non-relativistic fermion velocity. 
The anisotropic fermion velocity introduces an additional loss of symmetry and complicates calculations. Furthermore, since both components of the fermion velocity are small (relative to the photon velocity), the effective coupling is large and non-perturbative methods are required. 
We  use a Schwinger-Dyson (SD) approach (see \cite{pan} for some discussion of the use of SD methods in the context of  3d semi-metals) and derive a set of integral equations that describe the non-perturbative behaviour of this system and solve them numerically. 
We use throughout the Coulomb approximation in the photon propagator, which is well justified from the fact that the photon velocity is much larger than that of the electrons. 
We will also use a one loop approximation for the photon polarization tensor, which is motivated by the vanishing of the fermion density of states at the Dirac points and is very commonly adopted in the literature. In Ref. \cite{xiao2} the authors make additional restrictive approximations for the photon polarization tensor that significantly reduce the numerical difficulty of the calculations. Two of these approximations are the instantaneous approximation, $\Pi_{00}(p_0,\vec p) \to \Pi_{00}(0,\vec p)$, and the Khveshchenko approximation \cite{khv}, $\Pi_{00}(p_0,\vec p) \to \Pi_{00}(|\vec p|,\vec p)$. 
Ignoring the frequency dependence of the polarization tensor drastically simplifies the numerical  procedure that must be implemented to solve the Schwinger-Dyson equations and reduces the computation time by roughly a factor of $10^3$ 
(depending on the number of grid points used and the details of the interpolation and integration procedures).
However, a calculation that does not correctly take into account the frequency dependence of the photonic degrees of freedom can produce qualitatively different results (see, for example, Ref. \cite{mec1}). 
We include full frequency effects and compare with the results obtained from the instanteous approximation. 
Our results show that the instantaneous approximation gives a critical coupling that is artificially large. 

We will also consider the dependence of the phase transition on the fermion flavour number $N_f$. 
The critical coupling will be smaller for smaller values of $N_f$, which give less screening of the Coulomb interactions. 
The physical flavour number is 2, corresponding to the two Dirac cones in the Brillouin zone, but consideration of other flavour numbers is a useful way to estimate the possible effects of different approximations to the photon polarization tensor.   
For example it is known that in 2d Dirac materials calculations that include full photon dynamics, including a back-coupled polarization tensor obtained from a self-consistently solved SD equation, exhibit less screening relative to the one loop approximation \cite{mec2}. Mathematically this could be modelled with a smaller value of $N_f$.

The computation of the critical coupling is numerically difficult for several reasons. One is the size of the momentum phase space involved in the calculation of the dressing functions. 
Another problem, which occurs with all numerical studies of this type, is commonly called `critical slowing down.' This refers to the behaviour of the iterative procedure that is used to find solutions to non-perturbative integral equations in the vicinity of a critical point. Convergence to a solution becomes more difficult when the solution that is sought is close to the trivial solution where all dressing functions are zero. It is particularly important to obtain an accurate result for the gap function when it is close to zero  because the critical coupling, which is the value of the coupling for which the gap goes to zero, is necessarily determined by extrapolation.
The numerical procedure we have implemented is briefly described in section \ref{general-sec} and more details are given in Appendix \ref{numerical-app}.
We emphasize that the numerics of Ref. \cite{xiao2} were not accurate enough to draw reliable conclusions about the behaviour of the critical coupling. 

We study the dependence of the phase transition on the fermion velocity $v_F$, the flavour number $N_f$, and anisotropy parameter $\eta$. For physical values of these parameters, our calculations  produce values of the critical coupling that are larger than those in currently discovered  materials. 
However, our results show that the critical coupling and the nature of its dependence on $\eta$ depend strongly on the type of approximation used for the photon polarization tensor.  Calculations that include full photon dynamics, including a back-coupled polarization tensor obtained from a self-consistently solved SD equation, are necessary to answer even the basic question of whether the critical coupling increases or decreases with anisotropy. 
We conclude that the important question of whether anisotropy could provide a mechanism to
promote dynamical gap generation in realistic  three-dimensional Dirac
semi-metallic materials is still open.

\section{Notation}

Throughout this paper we work in Euclidean space. Four vectors are denoted with capital letters, for example $P_\mu = (p_0,\vec p)$, $P^2=p_0^2+p^2$, and 
we define $Q=K-P$. 
We use natural units: $c=\hbar=1$. 

\subsection{Feynman Rules}

To define the bare Feynman rules we introduce the matrix 
\bea
\label{Mdef}
M = 
\left[\begin{array}{cccc}
~1~ & ~0~ & ~0~ &0\\
0 & v_1  & 0 &0\\
0 & 0 & v_1  &0  \\
0 & 0 & 0  &v_3  \\
\end{array}
\right]\,.
\eea
The bare fermion propagator and bare vertex are
\bea
\label{bareFR}
&& S^{(0)}(p_0,\vec p) = -\big[i\gamma_\mu M_{\mu\nu} P_\nu\big]^{-1}\,\\[2mm]
\label{barevert}
&& \Gamma^{(0)}_\mu = M_{\mu\nu}\gamma_\nu\,.
\eea
We work in Lorentz gauge where the bare photon propagator has the form
\bea
&& G^{(0)}_{\mu\nu}(p_0,\vec p)=\big[\delta_{\mu\nu}-\frac{P_\mu P_\nu}{P^2}\big]\,\frac{1}{P^2}\, .
\eea
The dressed propagators are
\bea
&& G_{\mu\nu}^{-1}(p_0,\vec p) = G^{(0)-1}_{\mu\nu}(p_0,\vec p)+\Pi_{\mu\nu}(p_0,\vec p)\,\nn\\[2mm]
\label{SF0}
&&  S^{-1}(p_0,\vec p) = S^{(0)-1}(p_0,\vec p)+\Sigma(p_0,\vec p) 
\eea
where $\Pi_{\mu\nu}(p_0,\vec p)$ is the photon polarization tensor and $\Sigma(p_0,\vec p)$ is the fermion self-energy.

In vacuum QED the fermion self-energy has two independent components. In our problem, because of the broken Lorentz invariance, we have four components. We define
\bea
\label{SF}
&&  S^{-1}(p_0,\vec p) = S^{(0)-1}(p_0,\vec p)+\Sigma(p_0,\vec p) = - i \gamma_\mu H_{\mu\tau}(p_0,\vec p) P_\tau + D(p_0,\vec p)\,
\eea
where 
\bea
\label{Amatrix}
H(p_0,\vec p) = 
\left[\begin{array}{cccc}
Z(p_0,\vec p) & 0 & 0 &0\\
0 & v_1 A_1(p_0,\vec p)  & 0&0 \\
0 & 0 & v_1 A_1(p_0,\vec p)   &0 \\
0 & 0 & 0   & v_3 A_3(p_0,\vec p) \\
\end{array}
\right]\,.
\eea
Inverting equation (\ref{SF}) we obtain the fermion propagator in the form
\bea
&& S(p_0,\vec p) = \left(i \gamma_\mu H_{\alpha\beta}(p_0,\vec p) P_\beta +D(p_0,\vec p) \right) \frac{1}{{\rm den}_p} \nonumber \\[2mm]
&& {\rm den}_p = p_0^2 [Z(p_0,\vec p)]^{2} + v_1^2(p_1^2+p_2^2) [A_1(p_0,\vec p)]^2 + v_3^2 p_3^2 [A_3(p_0,\vec p)]^2 + D\big[(p_0,\vec p)\big]^2 \,. \label{prop-anio}
\eea
The dressing function $D(p_0,p)$ can give rise to a non-zero condensate $D(0,0)$ that plays the role of an effective mass. As discussed in section \ref{intro-sec}, a condensate is expected to provide an order parameter for a possible transition to an insulating state.

\subsection{Integral equations}

The SD equation for the fermion self energy is
\bea
\label{fermion-SD}
&& \Sigma(p_0,\vec p) = e^2\int dK \,G_{\mu\nu}(q_0,\vec q)\,M_{\mu\tau}\,\gamma_\tau \, S(k_0,\vec k) \,\Gamma_\nu\,
\eea
where $\Gamma_\nu$ denotes a vertex function. In principal this vertex should be obtained from its own SD equation, which in turn depends on a non-perturbative four-vertex. To truncate the hierarchy of SD equations, and to avoid the necessity of solving vertex integral equations, we will use an ansatz for the vertex $\Gamma_\mu$. The choice of this ansatz is discussed below. 
The projection operators for the four anisotropic fermion dressing functions are
\bea
\{ P_Z,P_{A_1},P_{A_3},P_D \} = \{\frac{i\gamma_0}{4p_0} , \frac{i(p_1\gamma_1+p_2\gamma_2)}{4v_1(p_1^2+p_2^2)}, \frac{i\gamma_3}{4v_3 p_3},\frac{1}{4} \}\,. \label{fermion-proj}
\eea

Since the electrons are non-relativistic we have $\{v_1,v_3\}\ll 1$. In this limit only the 00 component of the photon propagator contributes to the fermion self-energy, and only the 00 component of the photon polarization tensor contributes to the photon propagator $G_{00}$. Combining we have that the only part of the photon propagator that we need is  
\bea
G_{00}(q_0,\vec q) = \frac{q^4}{Q^4(q^2+\Pi_{00}\big(q_0,\vec q)\big)}\,.
\label{propY}
\eea
We will use throughout the Coulomb approximation which means we take $Q^2 \to q^2$ in equation (\ref{propY}). 
We will also make a common simplification in which $\Pi_{00}$ is not determined self-consistently but is obtained from a one loop calculation. 
The calculation of the one loop polarization tensor for this theory is given in Appendix \ref{appendixA}. The result is
\bea
&& \Pi_{00}(p_0,\vec p) = 4\pi \alpha v_1 (\text{part}_1 + \text{part}_2)\nonumber \\
&& \text{part}_1 = \frac{N_f }{12 \pi ^2 v_1^2 v_3} 
\tilde p^2 \, \text{Ln} \left(\frac{4\Lambda^2+\tilde P^2}{\tilde P^2} \right) \nonumber \\
&& \text{part}_2 = \frac{\Lambda^2 N_f }{3 \pi ^2 (\tilde P^2)^{3/2} v_1^2 v_3}
   \tilde p^2
   \left(2 \Lambda \cot ^{-1}\left(\frac{2
   \Lambda}{\sqrt{\tilde P^2}}\right)-\sqrt{\tilde P^2}\right) 
\label{us0}
\eea
where $\tilde P^2 = p_0^2+ \tilde p^2$ and $\tilde p^2 = v_1^2(p_1^2+p_2^2)+v_3^2p_3^2$. 
The parameter $\Lambda$ is an ultra-violet cutoff that is needed to 
define the polarization tensor. Since dynamical gap generation is a low-energy phenomenon we expect that the physics is dominated by 
the small momentum regime and therefore essentially independent of $\Lambda$. It is easy to verify numerically that this is true (see Appendix \ref{appendixA} for further discussion). 
The authors of \cite{xiao2} did not use equation (\ref{us0}) but instead used the simpler form 
\bea
\Pi_{00}^{\rm approx} = \frac{N_f}{6\pi^2 v_1^2 v_3}\tilde p^2 \, \text{Ln}\left(1+\frac{(v_1^2 v_3)^{1/3}\Lambda}{\sqrt{\tilde P^2}}\right)\,.
\label{his0}
\eea
They argued that the low momentum behaviour of the two expressions is similar enough that they should produce results for the critical coupling that are very close to each other. In figure \ref{pi-ratio} we look at the momentum behaviour of the two expressions  and in section \ref{results-sec} we make some comparisons of the critical couplings they produce.  The full expression for the polarization tensor is bigger than the approximation adopted by the authors of \cite{xiao2}, so the screening effect associated with it is larger, and the critical coupling it produces is therefore also larger. 

We begin by using the instantaneous approximation to the polarization tensor, bare vertices, and frequency independent dressing functions. This is the calculation of Ref. \cite{xiao2}. 
More physical results can be obtained using the full one loop polarization tensor, retaining the frequency dependence of the dressing functions, and introducing a suitable  ansatz for the vertex $\Gamma_\mu$
(as explained below). 
In both cases we use a shorthand notation that replaces the momentum arguments of the fermion dressing functions with a single subscript. In the instantaneous calculation $D_p$ indicates $D(\vec p)$ (and similarly for the other dressing functions). 
In the one loop calculation, $D_p$ means $D(p_0,\vec p)$, etc.
In all cases the distinction is clear from the form of the integral equation: in equations (\ref{3danio}, \ref{3diso}) the integral is over a three dimensional momentum space and all dressing functions are assumed to depend only on the spatial momentum; in equations (\ref{4danio}, \ref{4diso}) the frequency dependence is included in the dressing functions and the four dimensional momentum integral includes an integral over the frequency $k_0$.

\subsubsection{The instantaneous calculation}

We use a bare vertex in equation (\ref{fermion-SD}) and since $\{v_1,v_3\}\ll 1$ we need only its zero component. 
To satisfy the Ward identity we set $Z=1$. 
The integral over $k_0$ can be done analytically if we assume that the dressing functions depend only on spatial momenta and take $\Pi_{00}(q_0,\vec q) \to \Pi_{00}(0,q)$. 
The integral equations obtained after doing the $k_0$ integrals are
\bea
&& A_{\text{1p}} = 1+ \frac{4\pi v_1\alpha}{2(p_1^2+p_2^2)}\int\frac{d^3k}{(2\pi)^3} \frac{A_{\text{1k}} \left(k_1 p_1+k_2 p_2\right)}{ (q^2+\Pi_{00}(\vec q))
   \sqrt{v_1^2(k_1^2 + k_2^2 )A_{\text{1k}}^2 +k_3^2 v_3^2 A_{\text{3k}}^2+D_k^2}} \nonumber \\
&& A_{\text{3p}} = 1+  \frac{4\pi v_1\alpha}{2p_3} \int\frac{d^3k}{(2\pi)^3} \frac{k_3 A_{\text{3k}}}{ (q^2+\Pi_{00}(\vec q))  \sqrt{v_1^2(k_1^2 + k_2^2 )A_{\text{1k}}^2 +k_3^2 v_3^2 A_{\text{3k}}^2+D_k^2}} \nonumber \\
&& D_p = \frac{4\pi v_1\alpha}{2}\int\frac{d^3k}{(2\pi)^3}   \frac{D_k}{ (q^2+\Pi_{00}(\vec q)) \sqrt{v_1^2(k_1^2 + k_2^2 )A_{\text{1k}}^2 +k_3^2 v_3^2 A_{\text{3k}}^2+D_k^2}}\,. \label{3danio}
   \eea
As explained under equation (\ref{his0}), the fermion dressing functions depend on the spatial momentum components and this is indicated with a single subscript. 

\subsubsection{The one loop calculation}

The one loop polarization tensor is a good approximation to the full polarization tensor because of the vanishing of the fermion density of states at the Dirac points. We should also retain the frequency dependence of the fermion dressing functions, which means that the bare vertex is not an appropriate choice. 
The Ball-Chiu vertex is a well known ansatz that is constructed to truncate the SD equations without violating  gauge invariance \cite{ball-chiu-1,ball-chiu-2}. 
The full ansatz contains terms that are difficult to calculate numerically because they behave like ($0/0 \to $ finite) along special curves in the momentum phase space. A common approximation \cite{bc_1_a,bc_1_b,bc_1_c,mec1} is to take the first part of the ansatz  which has the form:
\bea
[\Gamma_\mu] = \frac{1}{2}\left(H_{\mu\alpha}(p_0,\vec p)+H_{\mu\alpha}(k_0,\vec k)\right)\gamma_\alpha\,.
\label{short}
\eea
The integral equations obtained using this vertex are
\bea
&& Z_p = 1- \frac{4\pi\alpha v_1}{2p_0}\int \frac{d^4k}{(2\pi)^4}\,\frac{k_0 Z_k(Z_k+Z_p)}{(q^2+\Pi_{00}(q_0,\vec q))(k_0^2 Z_k^2+v_1^2(k_1^2 + k_2^2 )A_{1k}^2 +k_3^2 v_3^2 A_{3k}^2+D_k^2)}\nonumber \\
&& A_{1p} = 1 + \frac{4\pi\alpha v_1}{2(p_1^2+p_2^2)}\int\frac{d^4k}{(2\pi)^4}\, \frac{A_{1k} (k_1p_1+k_2p_2) (Z_k+Z_p)}{(q^2+\Pi_{00}(q_0,\vec q))(k_0^2 Z_k^2+v_1^2(k_1^2 + k_2^2 )A_{1k}^2 +k_3^2 v_3^2 A_{3k}^2+D_k^2)} \nonumber \\
&& A_{3p} = 1 + \frac{4\pi\alpha v_1}{2p_3}\int\frac{d^4k}{(2\pi)^4}  \, \frac{k_3 A_{3k}(Z_k+Z_p)}{\big(q^2+\Pi_{00}(q_0,\vec q)\big)\big(k_0^2 Z_k^2+v_1^2(k_1^2 + k_2^2 )A_{1k}^2 +k_3^2 v_3^2 A_{3k}^2+D_k^2\big)}\nonumber \\
&& D_p = \frac{4\pi\alpha v_1}{2}\int\frac{d^4k}{(2\pi)^4}\,\frac{D_k(Z_k+Z_p)}{(q^2+\Pi_{00}(q_0,\vec q))(k_0^2 Z_k^2+v_1^2(k_1^2 + k_2^2 )A_{1k}^2 +k_3^2 v_3^2 A_{3k}^2+D_k^2)}\,. \label{4danio}
\eea
The fermion dressing functions in this equation depend on frequencies and the spatial momentum components. 

\subsection{Isotropic limit}
\label{iso-sec}

In the isotropic limit $v_1=v_3$ we should find $A_1=A_3$.
However, it is immediately clear that equations (\ref{us0}, \ref{3danio}, \ref{4danio}) do not give this result: if we set $v_1=v_3$ and $A_{1k}=A_{3k}$ on the right side, the functions $A_{1p}$ and $A_{3p}$ will clearly not be equal to each other. 
This happens because the propagator does not depend linearly on the dressing functions.\footnote{A solution to this problem in 1+2 dimensions can be found in \cite{brett1}.}

The correct way to do the isotropic calculation is to start from an isotropic decomposition of the fermion propagator by setting $A_1$ equal to $A_3$ in equation (\ref{Amatrix}) and using the projection operators
\bea
\{ P_Z,P_{A_1},P_D \}_{\rm iso} = \{\frac{i\gamma_0}{4p_0} , \frac{i(p_1\gamma_1+p_2\gamma_2+p_3\gamma_3)}{4v_1(p_1^2+p_2^2+p_3^2)},\frac{1}{4} \}\,. \label{fermion-proj-iso}
\eea
For bare vertices this gives
\bea
&& A_{\text{1p}} = 1+ \frac{4\pi v_1\alpha}{2(p_1^2+p_2^2+p_3^2)}\int\frac{d^3k}{(2\pi)^3} \frac{A_{\text{1k}} \left(k_1 p_1+k_2 p_2 + k_3 p_3\right)}{ (q^2+\Pi_{00}(\vec q))
\sqrt{v_1^2(k_1^2 + k_2^2  + k_3^2 )A_{\text{1k}}^2  + D_k^2}} \nonumber \\
&& D_p = \frac{4\pi v_1\alpha}{2}\int\frac{d^3k}{(2\pi)^3}   \frac{D_k}{ (q^2+\Pi_{00}(\vec q))
\sqrt{v_1^2(k_1^2 + k_2^2  + k_3^2 )A_{\text{1k}}^2  + D_k^2}}. \label{3diso}
\eea
Using the vertex ansatz in equation (\ref{short}) with $A_1=A_3$ and the projection operators in (\ref{fermion-proj-iso}) we obtain
\bea
&& Z_p = 1- \frac{4\pi\alpha v_1}{2p_0}\int \frac{d^4k}{(2\pi)^4}\,\frac{k_0 Z_k(Z_k+Z_p)}
{(q^2+\Pi_{00}(q_0,\vec q))(k_0^2 Z_k^2+v_1^2(k_1^2 + k_2^2 +k_3^2)A_{\text{1k}}^2 +D_k^2)}\nonumber \\[2mm]
&& A_{1p} = 1 + \frac{4\pi\alpha v_1}{2(p_1^2+p_2^2+p_3^2)}\int \frac{d^4k}{(2\pi)^4}\, \frac{A_{1k} (k_1p_1+k_2p_2 + k_3p_3) (Z_k+Z_p)}
{(q^2+\Pi_{00}(q_0,\vec q))(k_0^2 Z_k^2+v_1^2(k_1^2 + k_2^2 +k_3^2)A_{\text{1k}}^2 +D_k^2)} \nonumber \\
&& D_p = \frac{4\pi\alpha v_1}{2}\int \frac{d^4k}{(2\pi)^4}\,\frac{D_k(Z_k+Z_p)}
{(q^2+\Pi_{00}(q_0,\vec q))(k_0^2 Z_k^2+v_1^2(k_1^2 + k_2^2 +k_3^2 )A_{\text{1k}}^2 +D_k^2)}\,. \label{4diso}
\eea

\subsection{Scaling}
\label{scaling-sec}

To reduce the size of the momentum phase space, the isotropic equations (\ref{3diso}, \ref{4diso}) can be solved in spherical coordinates, and the anisotropic equations 
equations (\ref{3danio}, \ref{4danio}) can be solved in cylindrical coordinates. We represent this schematically as
\bea
&& \int d^3 k \,f_{\rm iso}(k_1,k_2,k_3) \to 4\pi \int_0^\Lambda dk\,k^2 \,\tilde f_{\rm iso}(k) \nn \\
&& \int d^4 k \,g_{\rm iso}(k_0,k_1,k_2,k_3) \to 4\pi \int_{-\Lambda}^\Lambda dk_0 \int_0^\Lambda dk\,k^2 \,\tilde g_{\rm iso}(k) \nn \\
&& \int d^3 k \,f_{\rm anio}(k_1,k_2,k_3) \to 2\pi \int_{-\Lambda}^{\Lambda} 
dk_3 \int_0^\Lambda dk\,k \,\tilde f_{\rm anio}(k,k_3)\,\nn \\
&& \int d^4 k \,g_{\rm anio}(k_0,k_1,k_2,k_3) \to 2\pi \int_{-\Lambda}^\Lambda dk_0 \int_{-\Lambda}^{\Lambda} 
dk_3 \int_0^\Lambda dk\,k \,\tilde g_{\rm anio}(k,k_3)\,.
\eea
The integral equations can then be written in a more useful form in terms of the dimensionless variables and functions:
\bea
&& p_1 \to \hat p_1 \Lambda  \text{~~ and likewise for all spatial components} \nn \\
&& p_0 \to v_1 \hat p_0 \Lambda \text{~~ and likewise for all frequencies } \nn \\
&& D_p \to v_1  \hat D_p \Lambda \text{~~ and likewise for }D_k \nn \\
&& \Pi_{00} \to \Lambda^2 \, \hat\Pi_{00}\,. \label{scale}
\eea
All factors of $\Lambda$ cancel after the integral equations are written in terms of the hatted quantities,  and it is therefore equivalent to replace all variables and functions with their hatted versions and adjust the upper limits of the momentum integrals to: $k\in(0,1)$, $k_3\in(-1,1)$ and $k_0\in(-1/v_1,1/v_1)$.  
We write the scaled polarization tensor in terms of the scaled variables
using the definitions $\eta=v_3/v_1$, $\bar P \equiv (\hat p_0^2+ \bar p^2)^{1/2}$ and $\bar p \equiv (\hat p_1^2 + \hat p_2^2 + \eta \hat p_3^2)^{1/2}$. The results are
\bea
&& \hat\Pi_{00} = \frac{\alpha N_f\bar p^2}{3\pi \eta v_1^3}\left(
\frac{8}{\bar P^3} \text{ArcCot}\left(\frac{2}{v_1\bar P }\right)
-\frac{4v_1}{\bar P^2}
+v_1^3  \text{Ln}\left(1+\frac{4}{v_1^2 \bar P^2}\right)
\right] \label{us2} \\[4mm]
&& \hat\Pi^{\rm approx}_{00} = \frac{2\alpha N_f \bar p^2}{3\pi\eta}\text{Ln}\left[1+\frac{\eta^{1/3}}{\bar P}\right] \,. 
\label{his2}
\eea
If we substitute the scaled version of the approximate polarization tensor (\ref{his2}) into the scaled versions of (\ref{3danio}, \ref{4danio}, \ref{3diso},  \ref{4diso}), and replace $v_3=\eta v_1$, all factors of $v_1$ cancel. Supressing the hats, it is equivalent to simply set $v_1=1$ in the integrand. In the instantaneous approximation, the $v_1$ dependence disappears completely. 
The full polarization tensor in equation (\ref{us2}) is explicitly dependent on $v_1$ even when it is written in terms of scaled variables. 

\begin{figure}[H]
\begin{center}
\includegraphics[scale=0.660]{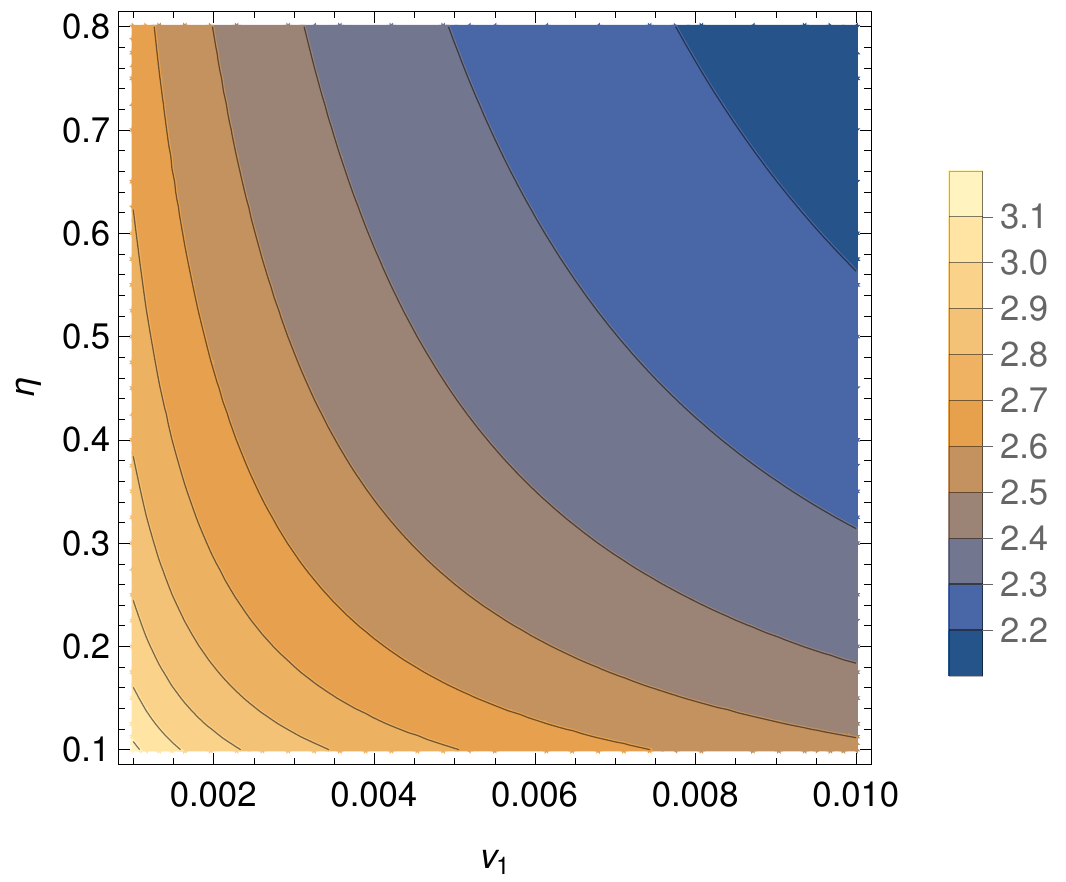} 
\includegraphics[scale=0.660]{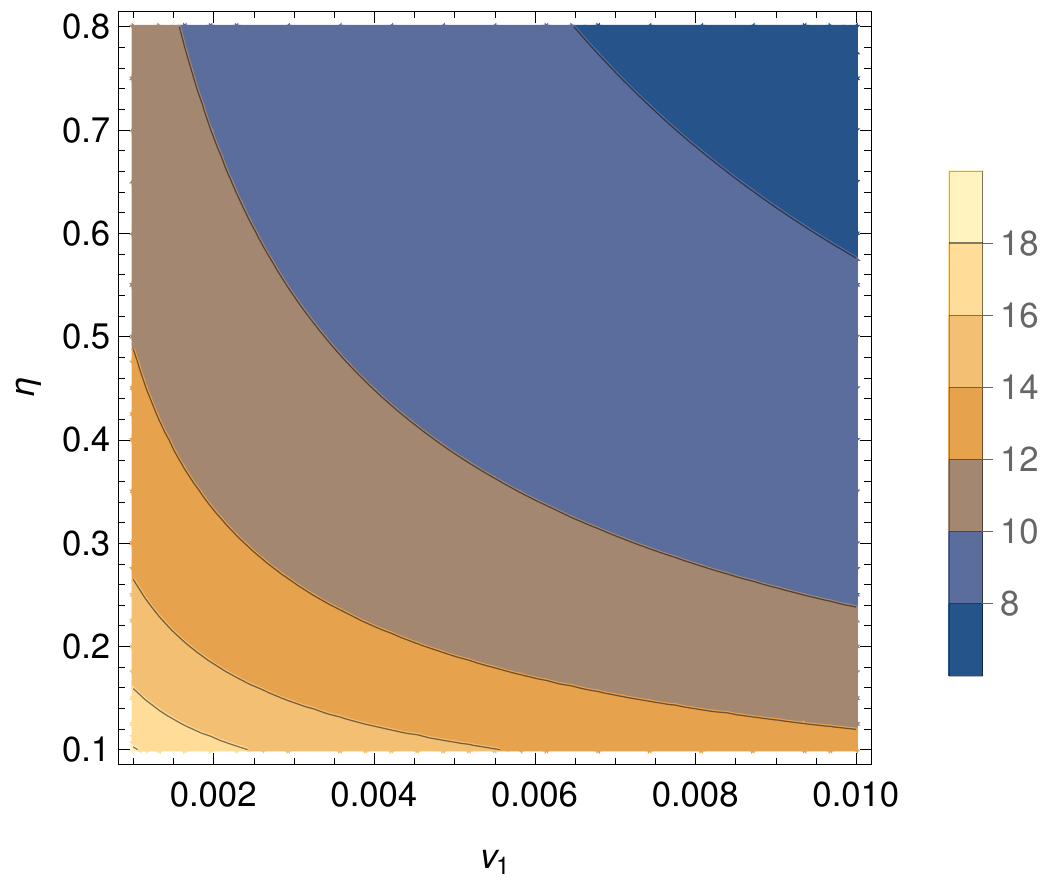} 
\caption{The ratio of the result in equation (\ref{us2}) divided by equation (\ref{his2}) at $\bar p = 0.1$ (left panel) and $\bar p=0.9$ (right panel). \label{pi-ratio}}
\end{center}
\end{figure}

As explained in Appendix \ref{appendixA}, the justification for using equation (\ref{his2}) is the idea that since dynamical gap generation is a low-energy
phenomenon, one expects that the dominant contribution to the integral equations for the fermion dressing
functions will come from the small momentum regime. To test the differences between the full and approximate polarization tensors, we show in figure \ref{pi-ratio} the ratio of equations (\ref{us2}, \ref{his2}) with $\hat p_0=0$ for a small value of $\bar p$ (left panel), and a large $\bar p$ (right panel), for different values of $\eta$ and $v_1$. 
The figures show that the correct expression is about 2 to 3 times larger at small momentum, and  10 to 15 times larger at large momentum. This indicates that the solutions of the integral equations for the fermion dressing function could change significantly when the approximate expression for the polarization tensor is used. 
In the next section we show some results that confirm this expectation. 

\section{Results}
\label{results-sec}

In this section we present some of our numerical results for the behaviour of the critical coupling of the metal-insulator transition in a Dirac 3d semi-metal. In all figures we use $v_1=0.01$, unless stated otherwise. In this section all variables are the dimensionless versions in (\ref{scale}) and we have suppressed the hats. 

\subsection{General considerations} 
\label{general-sec}

Our main goal is to study the effect of the parameter $\eta$, which quantifies the anisotropy of the perpendicular and parallel components of the fermion velocity. 
The authors of Refs. \cite{xiao2,xiao1} suggest that the natural definition $\alpha=e^2/(4\pi v_1)$ might not be a good choice when the purpose of the calculation is to study  anisotropy, and they propose that a better definition is 
 $\tilde\alpha =e^2/(4\pi \bar v)$ with $\bar v =(v_1^2 v_3)^{1/3}$ or $\tilde\alpha= \alpha/\eta^{1/3}$. 
The motivation for this choice was discussed in section \ref{intro-sec}.
We show below results from our calculation using both definitions. 

We will also consider different values of the flavour number $N_f$. 
The critical number of flavours is the value of $N_f$ for which the critical coupling goes to infinity. 
The critical coupling will be larger for larger values of $N_f$, which correspond to more screening of the Coulomb interactions. 
As explained in section \ref{intro-sec}, different values of $N_f$ are interesting as a way of estimating the effects of different approximations for the photon polarization tensor. 

The basic numerical procedure is as follows. For a chosen set of values for $(v_1,N_f,\eta,\alpha)$ we numerically solve one set of integral equations (\ref{3danio}, \ref{4danio}, \ref{3diso}, \ref{4diso}) using either (\ref{us2}) or (\ref{his2}) for the polarization tensor. 
The coupling is then reduced (at fixed $(v_1,N_f,\eta)$) to identify the value for which the condensate goes to zero, which is the critical coupling we are looking for. 
The same procedure is used for different values of $(v_1,N_f,\eta)$ to study the dependence of the critical coupling on these parameters. 
In the vicinity of the critical point the convergence procedure slows dramatically, because the solution that is sought is close to the trivial solution where all dressing functions are zero. However, it is particularly important to obtain  an accurate result for the gap function when it is close to zero, because the critical coupling itself  is necessarily determined by extrapolation. It is therefore absolutely essential to develop an accurate and efficient numerical procedure to perform these calculations. Some details of our method are given in Appendix \ref{numerical-app}.

\subsection{Instantaneous approximation and isotropic limit} 
\label{iso-sec-results}

\begin{figure}[H]
\begin{center}
\includegraphics[scale=1.00]{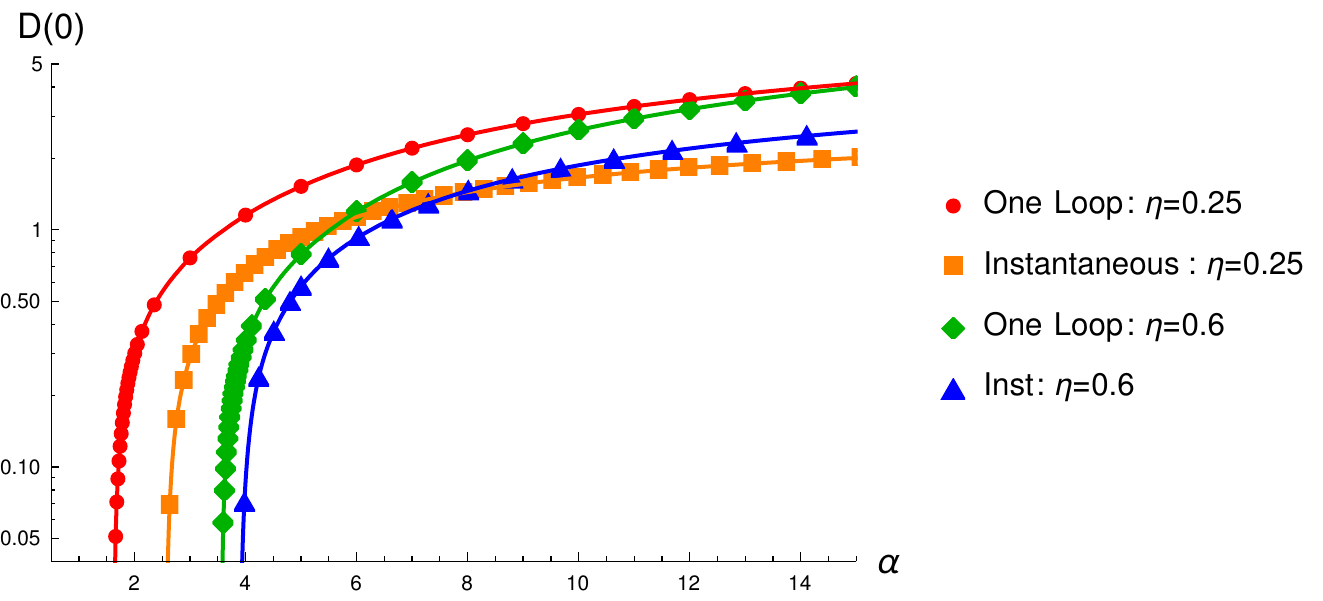} 
\end{center}
\caption{The condensate versus the coupling for $N_f=0.4$ and $v_1=0.01$ for two different values of $\eta$, using the instantaneous approximation and the full one loop polarization tensor.  \label{fig-inst}}
\end{figure}

The instantaneous approximation, which ignores the frequency dependence of the polarization tensor and the fermion dressing functions, is commonly used because it greatly reduces the numerical difficulty of the calculation. However, it is known that in some cases this approximation can have a qualitative effect on the critical coupling. For example,  in a study of graphene using a low energy effective theory approach, the instantaneous approximation gives a critical coupling that is too large by a factor of 2-3 \cite{mec1}. 
In figure \ref{fig-inst} we show the condensate versus the coupling for two different values of the anisotropy parameter, for $N_f=0.4$ and $v_1=0.01$, using the instantaneous approximation and the full one loop polarization tensor. The figure shows that the instantaneous approximation  produces an artificially large value for the critical coupling. These results motivate a detailed study that includes frequency dependent dressing functions of the dependence of the critical coupling on anisotropy and fermion velocity. 

\begin{figure}[H]
\begin{center}
\includegraphics[scale=1.00]{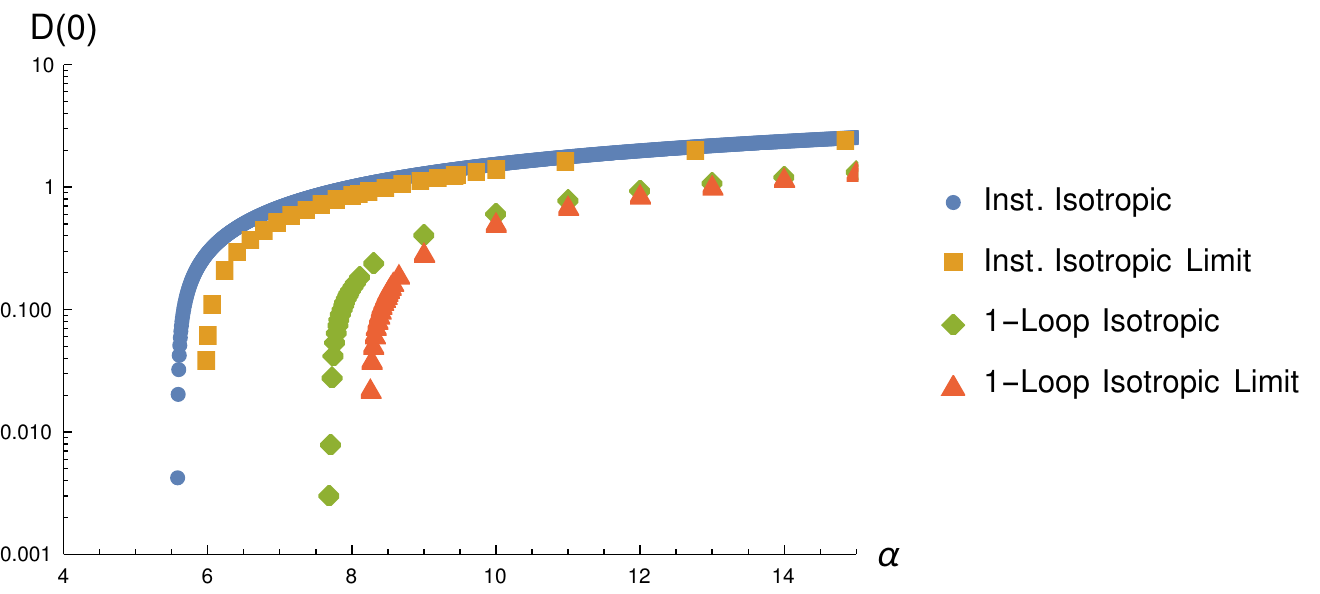} 
\end{center}
\caption{The condensate versus the coupling using the correct isotropic equations (\ref{3diso}, \ref{4diso}), and the isotropic limit of the anisotropic equations (\ref{3danio}, \ref{4danio}), for $N_f=0.1$. See text for further discussion.  \label{fig-iso}}
\end{figure}

In section \ref{iso-sec} we discussed the fact that the isotropic limit of the anisotropic calculation and the fully isotropic formalism do not completely agree with each other. 
Since we want to study the dependence of the critical coupling on anisotropy, it is important to understand to what extent the anisotropic results are reliable when the anisotropy parameter is close to the isotropic limit $\eta=1$. 
In figure \ref{fig-iso} we show the condensate versus the coupling using the correct isotropic equations (\ref{3diso}, \ref{4diso}), and the isotropic limit of the anisotropic equations (\ref{3danio}, \ref{4danio}). 
The figure indicates the critical coupling obtained from the anisotropic formalism with $\eta\lesssim 1$ will be slightly larger than the correct value.  
We will return to this point in section \ref{eta-sec} where  we study the dependence of $\alpha_c$ on $\eta$. 

\subsection{Velocity dependence} 
\label{v1-sec-results}

\begin{figure}[H]
\begin{center}
\includegraphics[scale=1.00]{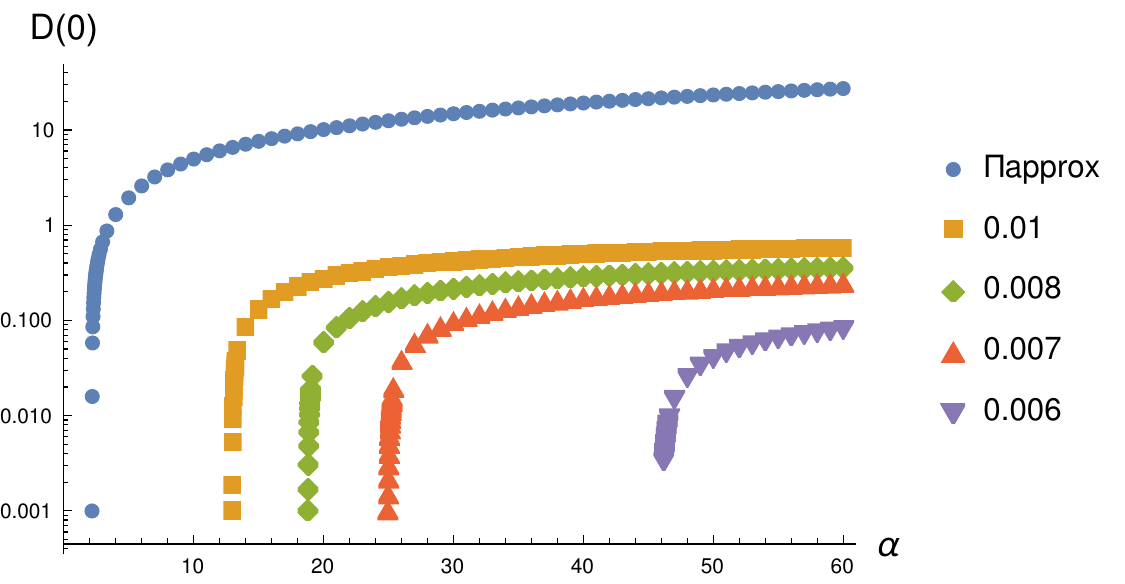} 
\end{center}
\caption{The condensate versus the coupling for $N_f=0.21$ and $\eta=0.4$ using the approximate result for the polarization tensor in equation (\ref{his2}), and the full expression in equation  (\ref{us2}) with four different values of $v_1$. \label{fig-v1}}
\end{figure}

In Ref. \cite{xiao2} it was suggested that it might be possible to promote dynamical gap production by finding a way to decrease $v_1$, or by finding new 3d Dirac semi-metals that have smaller values of the fermion velocity. The idea is that since $\alpha \sim e^2/v_1$, smaller $v_1$ corresponds to stronger interactions that would tend to reduce the critical coupling. 
However, this simple argument is valid only when one uses the approximate expression for the polarization tensor in equation (\ref{his2}). The correct polarization tensor in equation (\ref{us2}) depends explicitly on the fermion velocity and calculations are needed to determine how the critical coupling depends on the fermion velocity. 
In figure \ref{fig-v1} we show the condensate versus the coupling for $N_f=0.21$ and $\eta=0.4$ for four different values of the fermion velocity. The figure shows that when the correct polarization tensor is used, smaller values of $v_1$ actually produce larger critical couplings. We also show the curve obtained from the approximate polarization tensor in equation (\ref{his2}), which produces an artificially small value for the critical coupling. For reference we note that in Cd$_3$As$_2$ and Na$_3$Bi the fermion velocities are approximately 0.004 \cite{Cd3As2,Cd3As2-b} and 0.001 \cite{Na3Bi}, respectively. 

\subsection{Anisotropy dependence}
\label{eta-sec}

In Ref. \cite{xiao2} it was predicted that smaller values of $\eta$ would decrease the critical coupling. 
This is interesting because it would mean that the possibility exists that very large anisotropies could produce a condensate at physical values of the coupling. 
We note that in Cd$_3$As$_2$ and Na$_3$Bi the anisotropy parameters are approximately 0.25 \cite{Cd3As2,Cd3As2-b} and 0.1 \cite{Na3Bi}, respectively. 
In this section we show our results for the dependence of the critical coupling on anistropy, obtained using frequency dependent fermion dressing functions, and the full one loop expression for the polarization tensor in equation (\ref{us2}) which correctly includes the dependence on both the photon frequency and the fermion velocity.  

\begin{figure}[H]
\begin{center}
\includegraphics[scale=1.10]{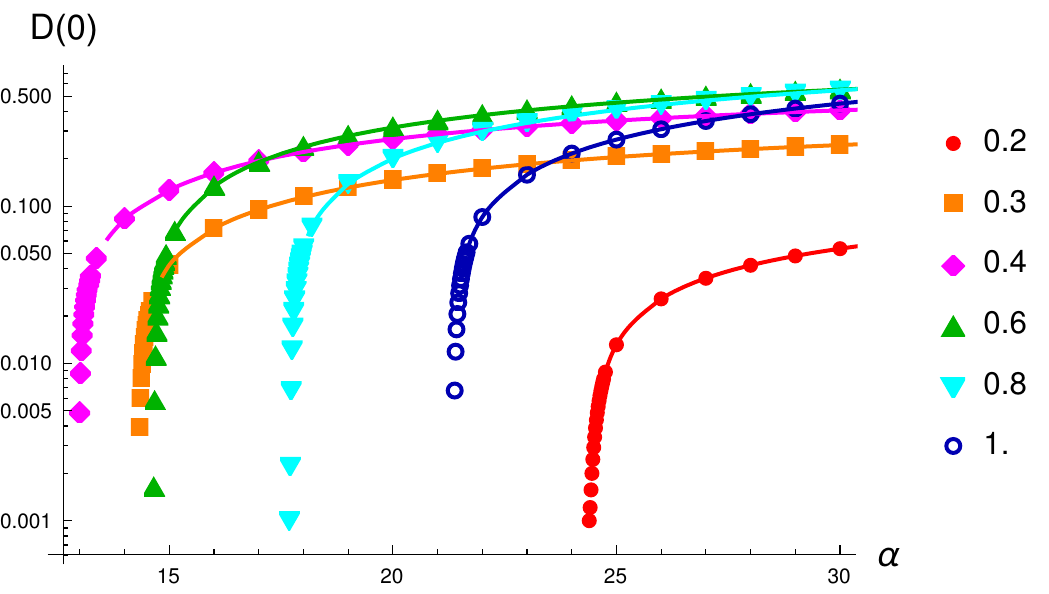} 
\end{center}
\caption{The condensate versus the coupling with $N_f=0.21$ using the full polarization tensor in equation (\ref{us2}), for different values of the anisotropy parameter $\eta$. 
\label{ploteta1}}
\end{figure}

In figure \ref{ploteta1} we show the condensate versus the coupling for $N_f=0.21$ for different values of the anisotropy parameter. The figure shows that for anisotropies $\eta\gtrsim 0.3$ the prediction of \cite{xiao2} is correct, but for smaller anistropies the dependence is in fact reversed, with smaller anisotropies producing larger critical couplings. 

\begin{figure}[H]
\begin{center}
\includegraphics[scale=0.80]{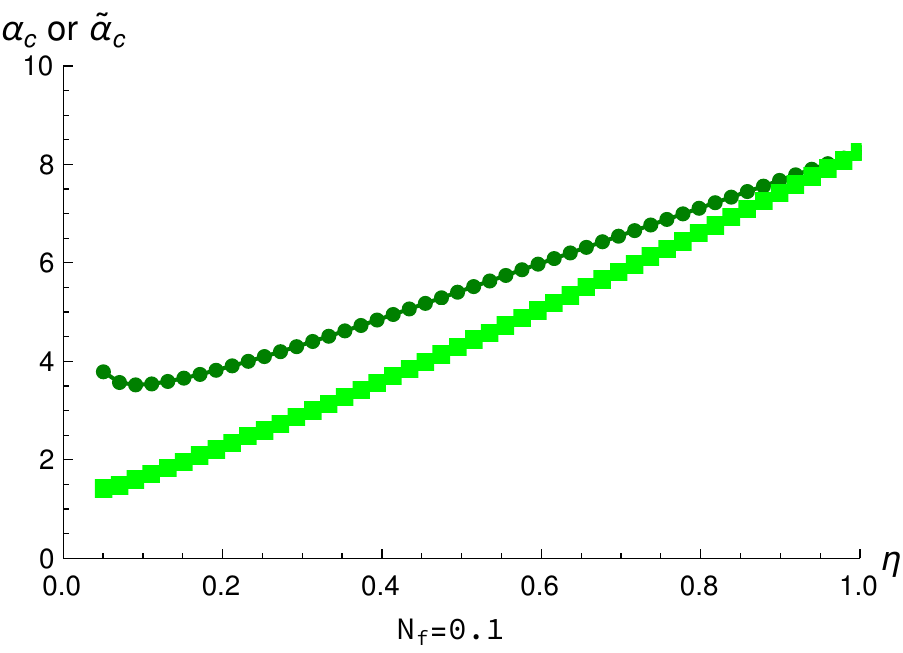} 
\includegraphics[scale=0.80]{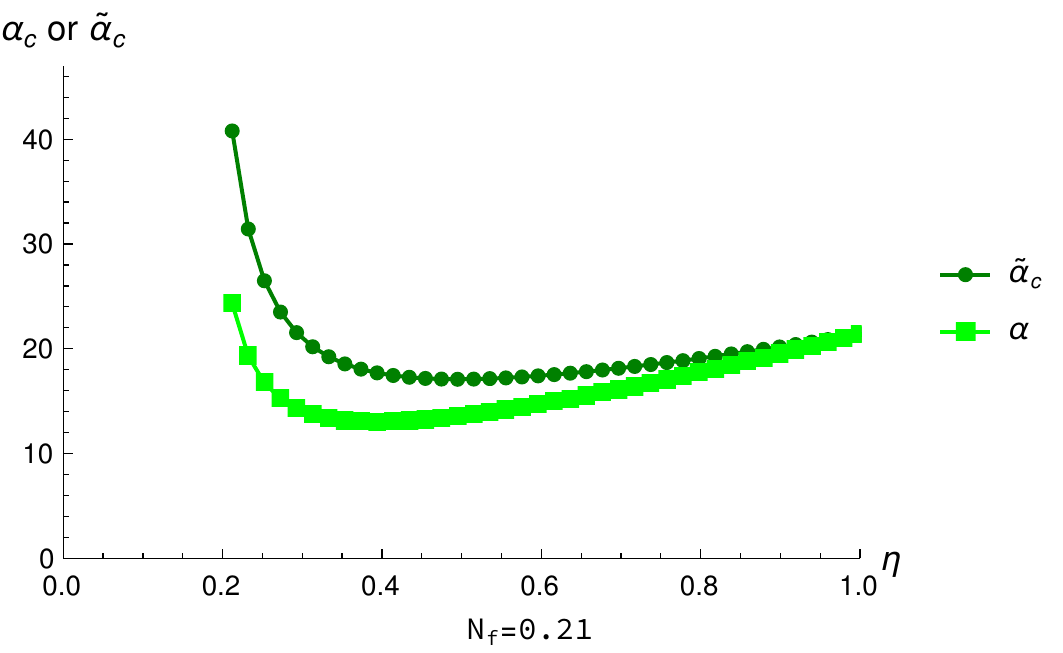} 
\end{center}
\caption{The critical coupling versus the anisotropy parameter $\eta$ for $N_f=0.1$ (left panel) and $N_f=0.21$ (right panel). \label{ploteta2}}
\end{figure}

In figure \ref{ploteta2} we show the $\eta$ dependence of the critical coupling, and the alternative definition of the critical coupling discussed in section \ref{general-sec}, for two different values of $N_f$.
The errors for each value of the critical coupling are calculated as explained in Appendix \ref{numerical-app} and are too small to be visible as error bars on the plot because of the wide range on the vertical axis. The largest error was obtained with $N_f=0.21$ and $\eta=0.21$ which gave $\tilde\alpha_c=40.78\pm 0.15$ and the minimum error was for $N_f=0.10$ and $\eta=0.29$ which gave $\alpha_c=2.8543 \pm 0.0007$. 
Since $\tilde\alpha_c = \alpha_c/\eta^{1/3}$ is largest relative to $\alpha_c$ at smallest values of $\eta$, the slope of  $\tilde\alpha_c$ versus $\eta$ changes the most relative to $\alpha_c$ versus $\eta$ at small $\eta$. The drop on the left side of the curves is therefore greater using the tilded definition of the critical coupling, but it is clearly not definition dependent. 
We also note that from the results in section \ref{iso-sec-results} we know that the formalism we are using slightly overestimates the critical coupling at large values of $\eta$, which means that the right end of the curves in both panels of figure \ref{ploteta2} are shifted slightly above the correct values. However, from figure \ref{fig-iso} it is clear that this effect does not change the fact that the right side of the curve has positive slope. We conclude that the dip in figure 6 is not an artifact of our calculation and the non-monotonic behaviour of the critical coupling as a function of $\eta$ indicates that there is complicated non-linear physics in play. 

\subsection{Critical flavour number}
\label{Nf-sec}

\begin{figure}[H]
\includegraphics[scale=1.10]{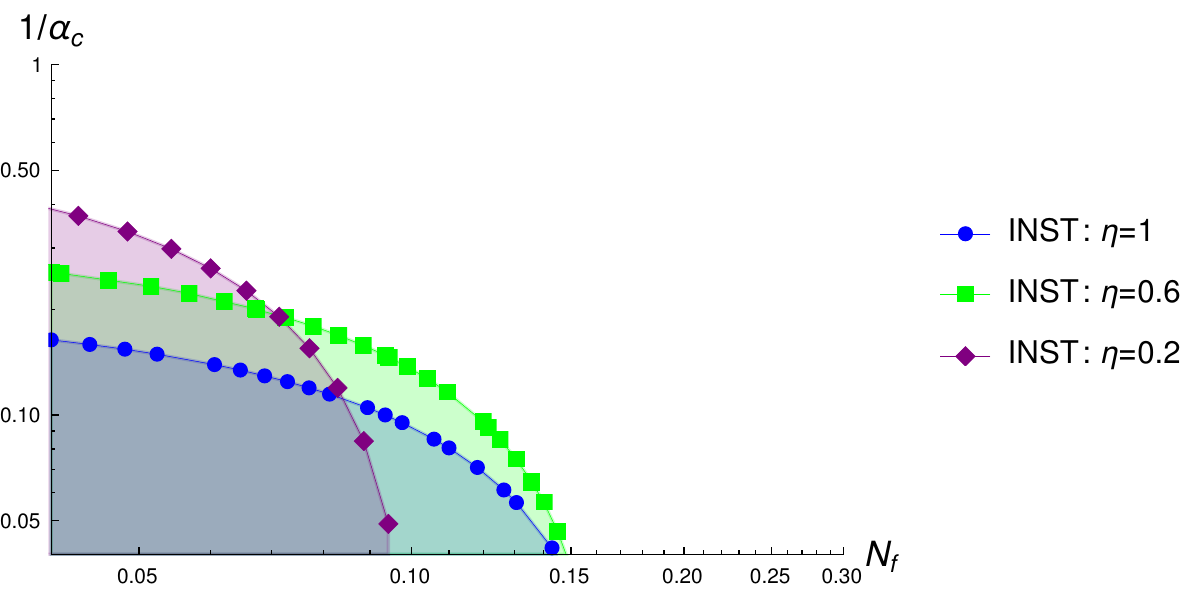} 
\includegraphics[scale=1.10]{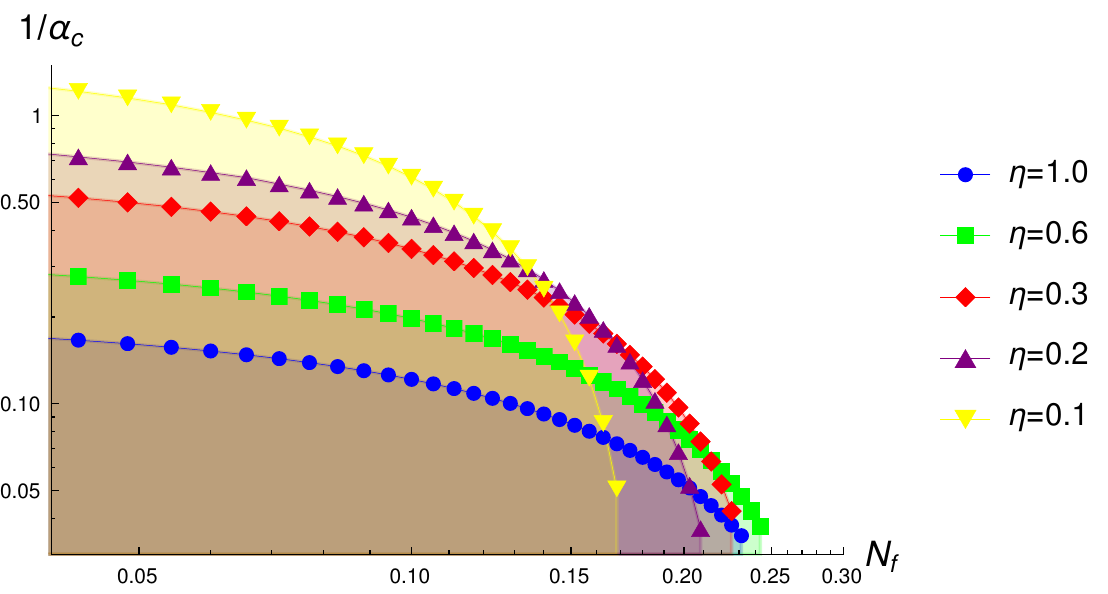} 
\caption{The shaded areas indicate the regions of $\alpha$ and $N_f$ where the condensate is non-zero. The top graph is the instantaneous approximation and the bottom graph is the one loop calculation. \label{plotPS3d}}
\end{figure}

One expects that smaller values of $N_f$ correspond to less screening, which promotes gap generation, and reduces the critical coupling. 
This behaviour is seen in figure \ref{plotPS3d} which shows the regions where the condensate is non-zero on a graph of $1/\alpha_c$ versus $N_f$, for both the instantaneous and one loop calculations. The insulating phase exists in the shaded regions. 
The data point with the largest error was $\eta=0.1$ and $N_f=0.02$ where the critical alpha was $\alpha_c = 1.475 \pm 0.018$.
The smallest error was for $\eta=1.0$ and $N_f=0.23$ which gave $\alpha_c=0.03484 \pm 0.00004$.
The crossing of the curves in the lower panel of figure \ref{plotPS3d} at $N_f=0.21$ is seen as the pronounced dip in the right panel of figure \ref{ploteta2}.
The curves in the figures can be used to obtain the critical flavour number, for which $1/\alpha_c$ goes to zero. The results are shown in table \ref{Nctable}. 

\begin{table}[H]
\begin{center}
\begin{tabular}{|c|c|c|}
\hline  
~~ $\eta$ ~~ & ~~~~~~ $N_{f{\rm crit}}$ inst ~~~~~~ & ~~~~ $N_{f{\rm crit}}$ 1-loop ~~~~   \\
 \hline
0.2 & 0.1021 $\pm$ 0.0005 & 0.224 $\pm$ 0.001 \\
0.6 & 0.168 $\pm$ 0.005 & 0.29 $\pm$ 0.01 \\
 \hline 
 \end{tabular}
\end{center}
\caption{The critical flavour number for two different values of the anisotropy parameter, using the instantaneous approximation and the full one loop polarization tensor. Results are obtained from the data used to produce figure \ref{plotPS3d}. \label{Nctable}}
\end{table}

\section{Conclusions}

We have studied the critical coupling of the semi-metal/insulator phase transtion in 3d Dirac semi-metals.  
We have used a non-perturbative Schwinger-Dyson approach and introduced an ansatz for the three-point vertex function to truncate the system of equations. 
The numerical calculations are difficult because the domain of the integrals is four dimensional, which means the number of loops/calculations is large. 
Another problem is that there are in general six independent dressing functions whose SD equations are coupled to each other, and relaxation methods typically struggle to find solutions when a large number of independent functions need to be simultaneously converged. In this work we have used a one loop approximation for the photon polarization tensor which, in the limit $v_F\ll1$, reduces the number of dressing functions to four. 
This is a commonly adopted approximation in the literature,  motivated by the vanishing fermion density of states at the Dirac points in 3d Dirac semi-metals. 

The main goal of this paper was to study the dependence of the critical coupling on the anisotropy parameter $\eta=v_3/v_1$ and the flavour number $N_f$. 
Our results for the critical coupling indicate that the critical flavour number is far below the physical value of 2 and therefore there should not be any observable evidence for insulating behavior, which is in general agreement with existing experiments. 

However, it is important to remember that the theory we are working with does not accurately treat short distance screening, in part because of the use of the one loop photon polarization tensor.
It is possible that more physical screening effects could be modelled using an ``effective'' flavour number that is smaller than the physical value.
This is supported by the results of Refs. \cite{mec2,brett2,brett3} where the critical coupling for the semi-metal/insulator transition in graphene was studied. In these papers it was found that photon backcoupling significantly decreased screening, which would correspond to a smaller value of $N_f$. 
It therefore seems reasonable to conclude that if the photon polarization tensor in our calculation were calculated self-consistently from its own SD equation, the behaviour of the critical coupling as a function of $\eta$ at the physical value $N_f=2$ might resemble the left panel of figure 6 (which shows a much smaller value of $N_f$). This means the result shown in the left panel of figure \ref{ploteta2},  that the critical coupling decreases fairly rapidly with increasing anisotropy at small values of $N_f$, is potentially important. 

The backcoupled anisotropic calculation is numerically more demanding than the calculations we presented in this paper, but is definitely feasible using the numerical techniques we have developed. These physically more realistic calculations would allow us to determine more decisively how the critical coupling is affected by anisotropy, and if a physically realizable critical coupling can be obtained. 

{\bf Data availability statement}

The data used to produce the figures is available on request. 

{\bf Author contribution statement}

All authors have contributed equally to the work reported on in this paper. 

\begin{acknowledgments}
We gratefully acknowledge helpful discussions with Andrew Frey. 
This work has been supported by the Natural Sciences and
Engineering Research Council of Canada Discovery Grant program 
 from grant 2017-00028.
All authors have contributed equally to the work reported on in this paper. 
The data used to produce the figures is available on request. 

\end{acknowledgments}

\appendix

\section{One loop polarization tensor}
\label{appendixA}
The zero-zero component of the one loop polarization tensor is obtained from the integral
\bea
\Pi_{00}(p_0,\vec p) = - N_f e^2 \int \frac{d^4k}{(2\pi)^4} \, {\rm Tr}\big(\gamma_0 S^{(0)}(k_0,\vec k) \gamma_0 S^{(0)}(q_0,\vec q)\big) \,
\eea
where the bare fermion propagator is given in equation (\ref{bareFR}).
After tracing over the Dirac indices we obtain
\bea
\Pi_{00}(p_0,\vec p) = - 4 N_f e^2 \int \frac{d^4k}{(2\pi)^4} \frac{k_0 q_0 -v_1^2(k_1 q_1+k_2 q_2) - v_3^2 k_3q_3}{(k_0^2+v_1^2(k_1^2+k_2^2)+v_3^2 k_3^2)(q_0^2+v_1^2(q_1^2+q_2^2)+v_3^2 q_3^2)}\,.
\label{eqpi1}
\eea
We rewrite the denominator by introducing a Feynman parameter using
\bea
\frac{1}{d_k d_q} = \int_0^1 dx \frac{1}{(d_k(1-x)+d_q x)^2}
\eea
where $d_k$ and $d_q$ indicate the two factors in the denominator of (\ref{eqpi1}). 
We then perform a shift of the integration variables using the definition $S_\mu = M_{\mu\nu}(K_\nu-x P_\nu)$. We define $ M^2 = x(1-x) (p_0^2+v_1^2(p_1^2+p_2^2)+v_3^2 p_3^2)$ and write the result
\bea
&& \Pi_{00}(p_0,\vec p) = \frac{ 4 N_f e^2 }{v_1^2 v_3} \int_0^1 dx\, \int \frac{d^4s}{(2\pi)^4}\frac{[s_0^2-s_1^2-s_2^2-s_3^2]-x(1-x)[p_0^2-p_1^2 v_1^2-p_2^2v_2^2-p_3^2v_3^2] }{s_0^2+s_1^2+s_2^2+s_3^2 + M^2} \,.\nn \\
\eea
We can now do the $s$ integrals in spherical coordinates. 
Introducing a cutoff $\Lambda$ on the momentum $|\vec s|$ we obtain the result in equation (\ref{us0}). 
We note that while a cutoff is needed to define the polarization tensor, it should not have a significant effect on the  critical coupling. 
The point is that since dynamical gap generation is a low-energy phenomenon, the dominant contribution to the integral equations for the fermion dressing functions should come from the small momentum regime. We have verified numerically that this is true. 

\section{Numerical method}
\label{numerical-app}

In this section we give some details of our numerical method. 

The integrals in equations (\ref{3danio}, \ref{4danio}, \ref{3diso}, \ref{4diso}), using either (\ref{us2}) or (\ref{his2}) for the polarization tensor, have an integrable singularity when $|\vec q| = |\vec k-\vec p| \to 0$. This means that the points in the domain of the integral where $\vec k \to \vec p$
must be treated carefully. We use Gauss-Legendre integration and divide the domain into regions so that all singular points are bracketed with equal sized intervals. This procedure allows for numerically accurate cancellations of large contributions on either side of the singularities. We also use a logarithmic scale for the momentum variables to increase sensititivity to the small momentum region where the dressing functions vary most strongly. 

To calculate the critical coupling, we start from a set of data points that gives the condensate $D(0)$ for different values of the coupling $\alpha$ (see figures \ref{fig-inst}, \ref{fig-iso}, \ref{fig-v1}, \ref{ploteta1}). 
To obtain this data, we start with a large value of $\alpha$ for which the condensate is not zero, and reduce the coupling step by step until the condensate goes to zero (numerically $10^{-4}$). 
We invert the array of data points to obtain a numerical representation of $\alpha[D(0)]$, construct an interpolated function, and extrapolate to find the critical coupling $\alpha_c\equiv \alpha[0]$. 
To obtain an accurate result it is clearly necessary to have a lot of data points for values of $\alpha$ that are close to the critical point, where the curve bends steeply downward. To do this in an efficient way, we dynamically adjust the step size as we approach the critical point. We start with the value $\Delta\alpha=1$ and, after the first two steps, the change in $\alpha$ is calculated from the results obtained from the previous two values as
\bea
\Delta\alpha = \left(\frac{1}{3}\frac{\alpha_{i-1}-\alpha_i}{\left(\ln[D(0)_{i-1}]-\ln[D(0)_i]\right)}\right)^2\,.
\eea 
This expression gives a large step size at large $\alpha$, when the value of the condensate changes slowly, and a smaller and smaller step size as the curve bends towards vertical.  
To prevent the step from becoming either too large or too small, we set a maximum step size of 1 and a minimum of 0.02. 

In all cases, the data that we need to interpolate is very smooth, and different interpolation methods give results for the critical coupling that agree to very high precision. 
It is clear however that the accuracy of the result for the critical coupling does not depend on the accuracy of the interpolated function. 
The smallest $\alpha$ points on the curves in figures (\ref{fig-inst}, \ref{fig-iso}, \ref{fig-v1}, \ref{ploteta1}) take the longest to calculate, but if the last five or six points were missing, the extrapolated critical coupling would be much too small. 
One way to quantify the error in the extrapolated result for the critical coupling would be to remove the last calculated point and compare with the previous result. It the data stopped at a point where the curvature of the data was large, this would give a significantly different critical coupling. However, if the last few points in the data give a line that is fairly straight but not close to vertical, this method would indicate a small error even though the extrapolated critical alpha will not be very accurate. 
An alternative estimate is the absolute difference of the extrapolated result and the smallest value of $\alpha$ in the data set. The error calculated this way is related to the inverse slope of the data at small $\alpha$, because it will be small if the curve drops steeply to the horizontal axis, and large if the curve is fairly flat. In all cases we have calculated the error both ways and taken the larger of the two results. 

\begin{figure}[H]
\begin{center}
\includegraphics[scale=0.80]{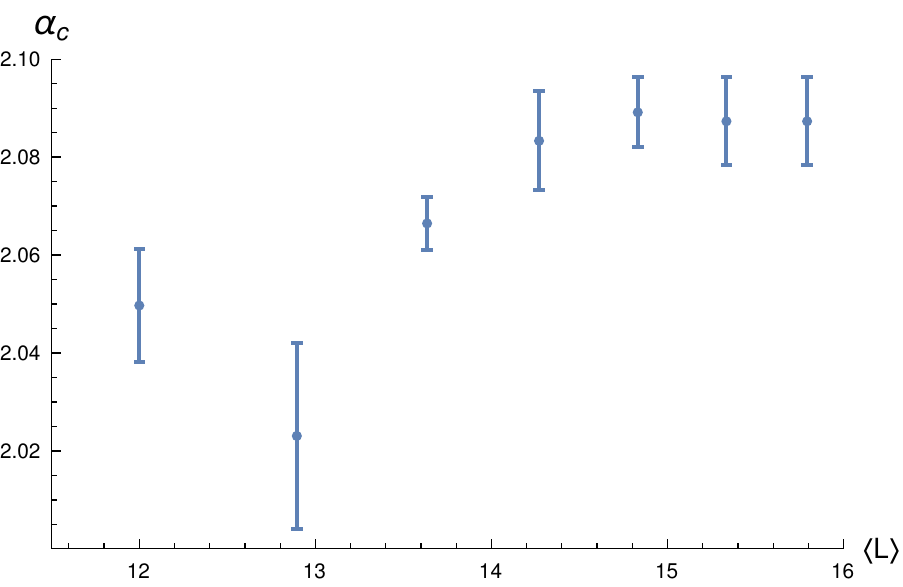} 
\end{center}
\caption{The critical coupling versus the average number of internal grid points per dimension. \label{grid-points}}
\end{figure}

The numerical solution of the SD equations involves four dimensional integrals and dressing functions that depend on four external variables. The total phase space therefore has a very large number of grid points, approximately $1.9 \times 10^9$. 
To verify that our integration and interpolation procedures are sufficiently accurate, we show in figure \ref{grid-points} a typical convergence plot. The graph shows the critical coupling, using the one loop polarization tensor for $\eta=0.4$ and $N_f=0.01$, as a function of the fourth root of the number of internal grid points (which gives the average value for one dimension). The graph indicates that good convergence is obtained for $\langle L \rangle \gtrsim 14$. In most of our calculations we used about 14.8. 

\end{document}